\documentclass[12pt,a4paper]{article}
\usepackage[utf8]{inputenc}
\usepackage{amsmath}
\usepackage{amsfonts}
\usepackage{amssymb}
\usepackage[left=2cm,right=2cm,top=2cm,bottom=2cm]{geometry}

\usepackage{graphicx}
\usepackage{booktabs}
\usepackage{url}

\usepackage{breakurl}
\usepackage[breaklinks]{hyperref}

\usepackage{authblk}
\usepackage{setspace}
\usepackage{float}

\usepackage{bm}

\title{\textbf{A comparison of approaches to incorporate patient-selected and patient-ranked outcomes in clinical trials}}

\author[a]{David S.\ Robertson\thanks{Corresponding author \texttt{david.robertson@mrc-bsu.cam.ac.uk}}}
\author[a,b]{Thomas Jaki}

\affil[a]{MRC Biostatistics Unit, University of Cambridge, UK}
\affil[b]{University of Regensburg, Germany}

\date{\vspace{-24pt}}

\begin{document}

\maketitle

\onehalfspacing

\begin{abstract}
A key aspect of patient-focused drug development is identifying and measuring outcomes that are important to patients in clinical trials. Many medical conditions affect multiple symptom domains, and a consensus approach to determine the relative importance of the associated multiple outcomes ignores the heterogeneity in individual patient preferences. Patient-selected outcomes offer one way to incorporate individual patient preferences, as proposed in recent regulatory guidance for the treatment for migraine, where each patient selects their most bothersome migraine-associated symptom in addition to pain. Patient-ranked outcomes have also recently been proposed, which go further and consider the full ranking of the relative importance of all the outcomes. This can be assessed using a composite DOOR (Desirability of Outcome Ranking) endpoint. In this paper, we compare the advantages and disadvantages of using patient-selected versus patient-ranked outcomes in the context of a two-arm randomised controlled trial for multiple sclerosis. We compare the power and type~I error rate by simulation, and discuss several other important considerations when using the two approaches.\\

\end{abstract}

\section{Introduction}
\label{sec:intro}

In recent years there has been a growing shift towards patient-focused drug development (PFDD), as exemplified by the series of guidance documents developed by the US Food and Drug Administration (FDA) from 2018--2023 \cite{FDApfdd}. A key part of PFDD is the identification of what aspects of symptoms and impacts of their condition are important to patients, so that clinical outcome assessments measure outcomes of importance to patients in clinical trials.

A challenge facing PFDD is that medical conditions can affect multiple symptom domains which are measured by multiple outcomes. This is particularly the case for complex disorders, where diseases such as multiple sclerosis (MS) or rheumatoid arthritis can affect people in diverse ways. Traditionally, stakeholders (including patients) come to a consensus around the relative importance of these different outcomes, resulting in the selection of (co-)primary and secondary endpoints~\cite{Mcleod2019}. However, such a consensus approach ignores the fact that improvement in some symptoms may be more important than others for each patient, as demonstrated by patient surveys in amyotrophic lateral sclerosis (ALS)~\cite{Van2022} and cardiovascular diseases~\cite{Chow2014}. It has been argued that a truly patient-centred approach should explicitly incorporate  heterogenous individual patient preferences into the evaluation of treatments~\cite{Janssens2019, Lu2022}.

A key step in using individual patient preferences in practice was the US FDA guidance for developing drugs for acute treatment for migraine, issued in 2018~\cite{FDA2018}. Migraine is a complex disorder with several symptoms, including pain, nausea, photophobia and phonophobia. In the past, the FDA required demonstration of an effect an all four of these symptoms (i.e.\ four co-primary endpoints). However, in the 2018 guidance, the FDA stated a preference for each patient being able to select the symptom (out of nausea, photophobia and phonophobia) that matters most to them:

\vspace{12pt}
\noindent ``\textit{A preferred approach, which aims to better align the study outcome with the symptom(s) of primary importance to patients, is to demonstrate an effect on both pain and the patient’s most bothersome symptom. Patients are asked to identify their most bothersome migraine-associated symptom in addition to pain. The identification can take place either before the attack is treated (e.g., at the baseline visit), or at the time of the attack, but before administration of the study drug. Using this approach, the two co-primary endpoints are (1) having no headache pain at 2 hours after dosing and (2) a demonstrated effect on the most bothersome migraine-associated symptom at 2 hours after dose}."
\vspace{12pt}

\noindent The guidance goes on to say that a binary outcome (present or absent) should be used for nausea, photophobia and phonophobia. Several trials have now been conducted using this `most bothersome symptom' endpoint~\cite{Munjal2020}. While there is now regulatory precedent in using such \textbf{patient-selected outcomes} as a (co-)primary endpoint, as far as we are aware, they have not been used outside of the context of migraine trials and binary data.

As an extension to the idea of using individual patient-selected outcomes, recent methodological work has proposed that instead of only evaluating efficacy based on the selected `top-ranked' outcome, individual patients can be asked to specify the full ranking of the relative importance of all relevant outcomes~\cite{Van2022, Lu2022}. These \textbf{patient-ranked outcomes} can be assessed using a composite DOOR (Desirability of Outcome Ranking) endpoint~\cite{Lu2022}, also known as PROOF (Patient-Ranked Order of Function) in the context of ALS~\cite{Van2022}. In this approach, every patient from one treatment arm is compared with every patient on another treatment arm, comparing whether each of the outcomes are more `favourable' while considering the difference in these patients’ ranking of outcome importance. The estimand is then the probability that, for a randomly selected pair of participants with one from the experimental arm and one from the control arm, the patient on the treatment arm has a more favourable composite DOOR than the patient on the control arm. 

In this paper, we take the trial setting used by Lu et al.~\cite{Lu2022} -- a two-arm randomised controlled trial (RCT) with continuous outcomes -- and compare the use of patient-ranked outcomes with patient-selected outcomes (i.e., only considering the top-ranked outcome). We propose two simple analysis methods for the patient-selected outcomes in this context, and compare them with composite DOOR, standard univariate analyses and an alternative analysis method for top-ranked outcomes proposed in~\cite{Lu2022}. Our goal is to compare the advantages and disadvantages of using the full patient ranking of all outcomes (patient-ranked outcomes) versus only using the top-ranked outcome (patient-selected outcomes) in this setting.

The rest of the paper proceeds as follows. In Section~\ref{sec:methods} we introduce the proposed outcomes and analysis methods for patient-selected outcomes, and recap the methodology for the patient-ranked outcomes. We present a simulation study in Section~\ref{sec:sim} based on a two-arm RCT for patients with MS, and conclude with a discussion in Section~\ref{sec:discuss}.

\section{Methods}
\label{sec:methods}

We focus on the setting of a two-arm RCT comparing an experimental treatment ($k = 1$) against a control ($k = 0$) with a total of $n = n_0 + n_1$ patients, where $n_0$ and $n_1$ denote the number of patients allocated to the control and experimental treatment, respectively. Let $\bm{Y_i} = (Y_{i,1}, \cdots, Y_{i,m})$ denote the vector of $m \geq 1$ outcomes for patient $i \in \{1, \ldots, n\}$ and $a_i \in \{0,1\}$ denote the treatment allocated to patient~$i$. For all analysis methods, the null hypothesis being tested is that there is no difference between the the experimental treatment and control, against a one-sided alternative hypothesis that the experimental treatment is `better' than the control, which we formalise for each outcome and analysis method below.

\subsection{Patient-selected outcomes}
\label{subsec:patient_selected}

Let $s_i \in \{1, \ldots, m\}$ denote the selected outcome of patient~$i$, i.e.\ the outcome that is most important or meaningful for that patient. The premise of patient-selected outcomes is that even if all outcomes $\bm{Y_i} = (Y_{i,1}, \cdots, Y_{i,m})$ are reported, what matters for the analysis is the single patient-selected outcome $Y_{i, s_i}$. Note that for a valid causal interpretation of the treatment effect, this selection mechanism should be independent of treatment assignment, which is satisfied when the selection cannot change and is determined prior to randomisation (see e.g.~\cite{Lu2022}). This prevents patient preference depending on their response to treatment.

\subsubsection*{Mean patient-selected outcome}

We assume that $Y_{i,j} \sim N(\mu_{a_i, j}, \sigma_{a_i}^2)$ independently for $j \in \{1, \ldots, m\}$, where $\sigma_{a_i}$ is unknown. We return to these assumptions below and in Section~\ref{sec:discuss}. A natural test statistic to test the null hypothesis that there is no difference in means between experimental arms (so that $H_0 : \mu_{0,j} = \mu_{1,j}$ for all $j \in \{1, \ldots, m\}$) is the Welch $t$-test, defined as follows:
\begin{equation}
t = \frac{\bar{Y}_1 - \bar{Y}_0}{\sqrt{d_0^2/n_0 + d_1^2/n_1}} \label{eq:t-statistic}
\end{equation}
where $\bar{Y}_k = \frac{1}{n_k} \sum_{i=1}^{n} 1\{ a_i = k\} Y_{i, s_i}$ and $d_k$ denotes the (corrected) sample standard deviation for $k \in \{0,1\}$, with $1\{\cdot\}$ denoting the indicator function.

An important subtlety with using this test statistic is ensuring that $\bar{Y}_1 - \bar{Y}_0$ has the required $N(0 , \frac{\sigma_0^2}{n_0} + \frac{\sigma_1^2}{n_1})$ distribution under the null. Under the null hypothesis we can write $\mu_{0,j} = \mu_{1,j} = \mu_j$. Letting $n_{k,j}$ denote the number of patients allocated to treatment~$k \in \{0,1\}$ who select outcome $j \in \{1, \ldots, m\}$, we have
\begin{align*}
\bar{Y}_1 - \bar{Y}_0 & \sim N\left( \frac{1}{n_1} \sum_{j=1}^m n_{1,j} \mu_j -  \frac{1}{n_0} \sum_{j=1}^m n_{0,j} \mu_j \, , \, \frac{1}{n_0^2}\sum_{j=1}^m  n_{0,j}\sigma_0^2 + \frac{1}{n_1^2}\sum_{j=1}^m  n_{1,j}\sigma_1^2 \right) \\
& = N\left( \sum_{j=1}^m \left( \frac{n_{1,j} }{n_1} - \frac{n_{0,j} }{n_0} \right) \! \mu_j \, , \, \frac{\sigma_0^2}{n_0} + \frac{\sigma_1^2}{n_1} \right)
\end{align*}
Hence two simple sufficient conditions are as follows:
\begin{enumerate}
\item $\frac{n_{1,j} }{n_1} = \frac{n_{0,j} }{n_0}$ for all $j \in \{1, \ldots, m\}$; or
\item $\mu_j = 0$ for all $j \in \{1, \ldots, m\}$.
\end{enumerate}

Condition~1 is satisfied when using stratified randomisation is such that $\frac{n_{1,j}}{n_1} = \frac{n_{0,j}}{n_0} $, i.e.\ stratified by the selected outcome. This stratified randomisation scheme would be recommended in practice in order to prevent (chance) imbalance between treatment groups for the known factor of selected outcome, as this could be prognostic of the outcome. For example, patients may rank their outcomes by baseline severity. Condition~2 may be appropriate if the control arm is a placebo. If either condition~1 or~2 are satisfied, then the test statistic given in equation~\eqref{eq:t-statistic} follows a $t$-distribution with degrees of freedom approximated by the Welch-Satterthwaite equation, see e.g.~\cite{Derrick2016}.

In general, we have \[
\bar{Y}_1 - \bar{Y}_0 \sim N\left(\sum_{j=1}^m \left( \frac{n_{1,j} }{n_1} \mu_{1,j} - \frac{n_{0,j} }{n_0} \mu_{0,j} \right) \, , \, \frac{\sigma_0^2}{n_0} + \frac{\sigma_1^2}{n_1} \right).
\]
Under condition~1 above, so that we can write $\frac{n_{1,j}}{n_1} = \frac{n_{0,j}}{n_0} = n'_j$, this simplifies to \[ 
\bar{Y}_1 - \bar{Y}_0 \sim N\left(\sum_{j=1}^m n'_j (\mu_{1,j} - \mu_{0,j}) \, , \, \frac{\sigma_0^2}{n_0} + \frac{\sigma_1^2}{n_1} \right).
\]
This motivates testing the null hypothesis against the one-sided alternative hypothesis $H_A : \mu_{1,j} \geq \mu_{0,j}$ for all $j \in \{1, \ldots, m\}$ (with a strict inequality for at least one~$j$), i.e.\ that the experimental treatment does not worsen any of the outcomes compared to the control.

\subsubsection*{Proportion test}

If the distributional assumptions on $Y_{i,j}$ do not hold, or neither of the two sufficient conditions above hold, then a simple alternative is to define a binary outcome for patient $i \in \{1, \ldots, n\}$ as follows:
\[Y_i^* = 1\{Y_{i, s_i} > \lambda_{s_i} \} \]
where $\lambda_{s_i}$ is some minimum clinically relevant effect size, such as the minimum clinically important difference (MCID), for outcome $s_i \in \{1, \ldots, m\}$. Let $p_k^* $ denote the unknown true proportion of patients who achieve a clinically meaningful improvement on their selected outcome, which can be  estimated by $\hat{p}_k^* = \sum_{i=1}^n 1\{a_i = k\} Y_i^* $.

The null hypothesis is then $H_0 : p_0^* = p_1^*$ versus the one-sided alternative $H_A: p_1^* > p_0^*$.  This can be tested using a Wald test to compare the proportion of patients achieving the minimum clinically relevant effect size on their selected outcome in the two treatment groups.

This outcome is similar in spirit to the outcome proposed in the FDA guidance on migraines, except that the underlying outcomes $Y_{i,j}$ are continuous. The dichotomisation would be expected to lead to a power loss, which we explore by simulation in Section~\ref{sec:sim}.

\subsection{Patient-ranked outcomes}

For the patient-ranked outcomes, we assume the outcome vector $\bm{Y_i}$ for each patient is paired with their importance ranking vector $\bm{R_i} = (R_{i,1}, \ldots, R_{i,m})$, where $R_{i,j} \in \{1, \ldots, m\}$ is the ranking that patient~$i$ assigns to outcome~$j$ with 1 being the most important and $m$ being the least important. Note that ties in the ranking are not allowed, and again for a valid causal interpretation of the treatment effect, this ranking mechanism should be independent of treatment assignment. For the patient-ranked outcomes described below, we only need to assume that $\bm{Y_i}$ are i.i.d from an arbitrary multivariate cumulative distribution function with location parameter $\bm{\mu_k}$ for treatment arm~$k \in \{0,1\}$.

\subsubsection*{Composite DOOR (Lu et al., 2022)}

Composite DOOR is built up from all pairwise comparisons of a univariate DOOR defined as follows: for a pair $(i_0, i_1)$ consisting of patient $i_0 \in \{1, \ldots, n_0\}$ from the control arm and $i_1 \in \{1, \ldots, n_1 \}$ from the experimental treatment, the DOOR indicator $Z_{i_0, i_1}(j)$ for the $j$th outcome is given by 
\[
Z_{i_0, i_1}(j) = \begin{cases}
1 \qquad \text{if} \quad Y_{i_1, j} \succ Y_{i_0, j}\\
0 \qquad \text{if} \quad Y_{i_1, j} \prec Y_{i_0, j} \\
0.5 \quad \; \text{if} \quad Y_{i_1, j} \bowtie Y_{i_0, j}
\end{cases}
\]
where $\succ$ and $\prec$ denote that the outcome in the experimental arm is more or less favourable than the control arm, respectively, while $\bowtie$ denotes that the two outcome are a tie.

In the rest of this paper, we assume that the $Y_{i,j}$ are continuous with the DOOR defined as below, following~\cite{Lu2022}:
\[
Z_{i_0, i_1}(j) = \begin{cases}
1 \qquad \text{if} \quad Y_{i_1, j} - Y_{i_0, j} > \text{MCID}_j\\
0 \qquad \text{if} \quad Y_{i_1, j} - Y_{i_0, j} < \text{MCID}_j\\
0.5 \quad \; \text{if} \quad |Y_{i_1, j} - Y_{i_0, j}| \leq \text{MCID}_j\\
\end{cases}
\]

The composite DOOR for all outcomes for a pair of patients $(i_0, i_1)$ is then derived following an iterative process that compares the set of outcomes that share the same set of rankings, starting from the smallest possible common set of rankings to the largest. For each common set~$S$, if $Y_{i_1, j} \succ Y_{i_0, j}$ for all $j \in S$ and there is no outcome $j' \in S$ such that $Y_{i_1, j'} \prec Y_{i_0, j'}$, then we conclude overall that $\bm{Y_{i_1}} \succ \bm{Y_{i_0}}$. Conversely, if $Y_{i_1, j} \prec Y_{i_0, j}$ for all $j \in S$ and there is no outcome $j' \in S$ such that $Y_{i_1, j'} \succ Y_{i_0, j'}$, then we conclude overall that $\bm{Y_{i_1}} \prec \bm{Y_{i_0}}$. If $Y_{i_1, j} \succ Y_{i_0, j}$ for at least one $j \in S$ and $Y_{i_1, j'} \prec Y_{i_0, j'}$ for at least one $j' \in S$, then we conclude overall that $\bm{Y_{i_1}} \bowtie \bm{Y_{i_0}}$. Finally, if all outcomes in~$S$ are tied (i.e.\ $Y_{i_1, j} \bowtie Y_{i_0, j}$) for all $j \in S$ then we examine outcomes in the next largest common set of rankings and so on. If the largest common set of rankings (that is, the set of all $m$ outcomes) is reached and all $m$ outcomes are still tied, then we set $\bm{Y_{i_1}} \bowtie \bm{Y_{i_0}}$. The composite DOOR, denoted $Z_{i_0, i_1}$, is then given by \[
Z_{i_0, i_1} = \begin{cases}
1 \qquad \text{if} \quad \bm{Y_{i_1} }\succ \bm{Y_{i_0}}\\
0 \qquad \text{if} \quad \bm{Y_{i_1}} \prec \bm{Y_{i_0}} \\
0.5 \quad \; \text{if} \quad \bm{Y_{i_1}} \bowtie \bm{Y_{i_0}}
\end{cases}
\]

The corresponding estimand is then the probability that, for a randomly selected pair of participants with one from the experimental arm and one from the control arm, the patient on the treatment arm has a more favourable composite DOOR than the patient on the control arm. More formally, the winning probability is defined as \[
\theta = P(\bm{Y_{i_1}}  \succ \bm{Y_{i_0}}) + \frac{1}{2}P(\bm{Y_{i_1}} \bowtie \bm{Y_{i_0}})
\]
where the second term is to account for ties. To estimate $\theta$, we use \[
\hat{\theta} = \frac{1}{n_0 n_1} \sum_{i_0 = 1}^{n_0} \sum_{i_1 = 1}^{n_1} Z_{i_0, i_1}
\]

The null hypothesis is $H_0 : \theta = 0.5$ versus the one-sided alternative $H_A : \theta > 0.5$ (note that a two-sided alternative was used in~\cite{Lu2022}). The null hypothesis is tested using an asymptotic normal approximation so that we reject the null hypothesis if $(\hat{\theta} - 0.5) / \hat{\sigma} \geq z_{1-\alpha}$ where $\hat{\sigma}$ is given in the Appendix (Section~\ref{Asec:DOOR}), $z_q$ denotes the $100 \times q\%$ percentile of the standard normal distribution and $\alpha$ is the (target) type~I error rate.
For further details of the composite DOOR algorithm and illustrations, we refer the reader to Lu et al.~\cite{Lu2022}. For convenience, we have reproduced the formal algorithm for composite DOOR in the Appendix (Section~\ref{Asec:DOOR}).

\subsubsection*{Top-ranked approaches}

Composite DOOR uses the full hierarchy of patient ranking and the associated outcomes. As a more direct comparator to the patient-selected outcomes described in Section~\ref{subsec:patient_selected}, one natural idea is to use a top-ranked version of composite DOOR, i.e.\ only calculate composite DOOR based on the top-ranked outcome of each patient. More precisely, for a a pair of patients $(i_0, i_1)$, if they share the same top-ranked outcome $j^*$ then $Z_{i_0, i_1} = Z_{i_0. i_1}(j^*)$, otherwise $Z_{i_0, i_1} = 0.5$. However, this will tend to result in a very large number of ties ($Z_{i_0, i_1} = 0.5$), leading to low power.

Instead, we use the top-ranked Weighted Winning Probability (WWP) approach, also proposed by Lu et al.~\cite{Lu2022}. This approach stratifies patients into groups based on which outcome they rank as top, and for each group calculates a separate winning probability. These are then combined by taking a weighted average of the group winning probabilities, where the weight for each group is how large that group is (i.e.\ the proportion of patients who ranked that outcome as top).

More formally, let $S_j$ denote the set of all patients who rank outcome $j$ as their top preference, i.e., $S_j = \{i : R_{i,j} = 1\}$. Also let $S_{0, j}$ and $S_{1, j}$ denote the subsets of $S_j$ for patients allocated to the control and treatment, respectively. For each stratum $j$, the winning probability $\hat{\theta}_j$ is calculated as follows.
\begin{equation*}
    \hat{\theta}_j = \frac{1}{n_{0,j} n_{1,j}} \sum_{i_0 \in S_{0,k}} \sum_{i_1 \in S_{1,k}} Z_{i_0, i_1}(j)
\end{equation*}

The weight for each stratum is the sample proportion of all patients who ranked outcome $j$ as their most important, denoted $\hat{p}_j$:
\begin{equation*}
    \hat{p}_j = \frac{|S_j|}{n} = \frac{1}{n} \sum_{i=1}^{n} 1\{R_{i,j} = 1\}
\end{equation*}

The overall weighted winning probability, $\hat{\theta}_{WT}$, is the weighted average of the within-stratum winning probabilities, using the stratum proportions as weights.
\begin{equation*}
    \hat{\theta}_{WT} = \sum_{j=1}^{m} \hat{p}_j \, \hat{\theta}_j
\end{equation*}

To test the null hypothesis $H_0: \theta_{WT} = 0.5$ against the one-sided alternative $H_A: \theta_{WT} > 0.5$, we use an asymptotic normal approximation so that the null hypothesis is rejected if $(\hat{\theta}_{WT} - 0.5)/\hat{\sigma}_{WT} \geq z_{1-\alpha}$ where $\hat{\sigma}_{WT}$ is given in the Appendix (Section~\ref{Asec:WWT}). 


\subsection{Standard analysis: univariate mean}

Finally, as a baseline comparison we include the standard analysis that ignores individual patient preferences and simply chooses \textit{a-priori} one of the outcome~$j^* \in \{1, \ldots, m\}$ as the primary outcome of interest for all patients. The null hypothesis is $H_0: \mu_{0, j^*} =  \mu_{1, j^*}$ versus the alternative $H_A : \mu_{1, j^*} >  \mu_{0, j^*}$. 

As a fair comparison to the $t$-test proposed in Section~\ref{subsec:patient_selected}, we assume that $Y_{i,j^*} \sim N(\mu_{a_i, j^*}, \sigma_{a_i}^2)$ independently for $i \in \{1, \ldots, n\}$, where $\sigma_{a_i}$ is unknown and use the Welch $t$-test to compare the means between the treatment and control groups on the outcome~$j^*$.

\subsection{Summary of methods}

Table~\ref{tab:method_summary} gives a high-level summary of the different outcomes and analysis methods described above. We build on this comparison of methods in the simulation study (Section~\ref{sec:sim}) as well as the Discussion (Section~\ref{sec:discuss}).

\begin{table}[ht!]
\def\arraystretch{1.5}
\footnotesize
\begin{tabular}{p{3.2cm} | p{6cm} p{2.8cm} p{2.8cm} } 
\textbf{Outcome/Method} & \textbf{Interpretation} & \textbf{Uses individual preference?} & \textbf{Information used} \\ 
\hline 
Composite DOOR & What is the overall probability that a random treated patient ``wins" against a random control patient, considering the full hierarchy of individual patient preference? & Yes (full hierarchy) & Full hierarchy of rankings and all outcomes  \\ 

Weighted top-ranked winning probabilities &  What is the average winning probability of the top-ranked outcome strata (weighted by proportion of patients in each strata)? & Yes (top-ranked) & Top-ranked outcome \\ 
Univariate mean & 	Is the mean of the each outcome (separately) different between treatment groups? & No & Single outcome \\ 
Mean patient-selected outcome & Is the mean of the patient-selected outcomes different between treatment groups? & Yes (patient-selected) & Selected outcome \\ 
Proportion test of patient-selected outcomes & Is the proportion of patients who achieve a clinically meaningful improvement on their selected outcome different between the treatment groups? & Yes (patient-selected) & Selected outcome\\  
\end{tabular}
\caption{Summary of patient-selected and patient-ranked outcomes.\label{tab:method_summary}}
\end{table}

\section{Simulation study}
\label{sec:sim}

For our simulation study, we use a two-arm RCT for patients with multiple sclerosis (MS) as described in Lu et al.~\cite{Lu2022}. The trial was planned to compare a cognitive and behaviour therapy (CBT) intervention against standard of care. Three clinical outcomes -- fatigue (outcome~1), pain (outcome~2) and depression (outcome~3) -- were measured using PROMIS scoring~\cite{PROMIS}. The primary outcomes were the normalised reduction in fatigue, pain and depression PROMIS scores from baseline after one year. The MCIDs equal 0.67, 0.63 and 0.54 for fatigue, pain and depression (respectively), as reported in Yost et al.~\cite{Yost2011}.

During the development of the trial protocol (i.e., prior to randomisation), a patient survey was conducted to elicit individual patient preferences regarding the relative importance of improvements on these three outcomes. The proportion of respondents that preferred each of the 6 possible rankings of importance were as follows:
\[ (p_{123}, p_{132}, p_{213}, p_{231}, p_{312}, p_{321}) = (0.42, 0.17, 0.24, 0.05, 0.08, 0.04 ). \]
Note this implies that 59\%, 29\% and 12\% of respondents ranked fatigue, pain and depression (respectively) as the most important outcome. In what follows, we make the natural assumption that the top-ranked outcome for each patient when using the patient-ranked methods is the same as their selected outcome when using patient-selected methods.

We now describe the simulation study using the framework of Morris et al.~\cite{Morris2019}. \\

\noindent \textbf{Aims}: to compare the use of patient-ranked and patient-selected endpoints as described in Section~\ref{sec:methods}, in order to offer a proof-of-concept of using patient-selected endpoints. \\

\noindent \textbf{Data-generating mechanisms}:
We follow Lu et al.~\cite{Lu2022} and assume that the three outcomes follow a multivariate normal distribution with covariance matrix \[
\Sigma = \left(\begin{smallmatrix}
1 & 0.55 & 0.55 \\
0.5 & 1 & 0.5 \\
0.55 & 0.5 & 1
\end{smallmatrix}\right)
\]

As a sensitivity analysis, we additionally consider a `low' correlation setting with covariance matrix \[
\Sigma = \left(\begin{smallmatrix}
1 & 0.25 & 0.25 \\
0.25 & 1 & 0.25 \\
0.25 & 0.25 & 1
\end{smallmatrix}\right)
\]
and a `high' correlation setting with covariance matrix \[
\Sigma = \left(\begin{smallmatrix}
1 & 0.75 & 0.75 \\
0.75 & 1 & 0.75 \\
0.75 & 0.75 & 1
\end{smallmatrix}\right).
\]

We consider a trial of 60 patients, and use 1:1 randomisation to CBT or usual care that is stratified by patient preference strata (with a randomisation block of size~2).
%
%
The mean reduction from baseline in the standard of care arm for the three outcomes is assumed to be $(0,0,0)$. Table~\ref{tab:sim_scenarios} shows the eight scenarios for the mean reduction from baseline in the CBT arm for the three outcomes, as given in~\cite{Lu2022}.

\begin{table}[ht!]
  \centering
  \footnotesize
  \renewcommand{\arraystretch}{1.5}
  \begin{tabular}{l c c c c c c c c c}
    \toprule
    & & \multicolumn{8}{c}{\textbf{Mean reduction from baseline for CBT}} \\
    \cmidrule(lr){3-10}
    \textbf{Ranking} & \textbf{Usual Care} & \textbf{S1} & \textbf{S2} & \textbf{S3} & \textbf{S4} & \textbf{S5} & \textbf{S6} & \textbf{S7} & \textbf{S8} \\
    \midrule
    (1,2,3) & (0,0,0) & (0,0,0) & (1,1,1) & (1,0,0) & (0,0,1) & (1,0,0.5) & (1,0.5,0) & (0,0.5,1) & (-1,1,0) \\
    (1,3,2) & (0,0,0) & (0,0,0) & (1,1,1) & (1,0,0) & (0,0,1) & (1,0,0.5) & (1,0,0.5) & (0,1,0.5) & (-1,1,0) \\
    (2,1,3) & (0,0,0) & (0,0,0) & (1,1,1) & (1,0,0) & (0,0,1) & (1,0,0.5) & (0.5,1,0) & (0.5,0,1) & (-1,1,0) \\
    (2,3,1) & (0,0,0) & (0,0,0) & (1,1,1) & (1,0,0) & (0,0,1) & (1,0,0.5) & (0,1,0.5) & (1,0,0.5) & (-1,1,0) \\
    (3,1,2) & (0,0,0) & (0,0,0) & (1,1,1) & (1,0,0) & (0,0,1) & (1,0,0.5) & (0.5,0,1) & (0.5,1,0) & (-1,1,0) \\
    (3,2,1) & (0,0,0) & (0,0,0) & (1,1,1) & (1,0,0) & (0,0,1) & (1,0,0.5) & (0,0.5,1) & (1,0.5,0) & (-1,1,0) \\
    \bottomrule
  \end{tabular}
\caption{{Simulation scenarios for the mean reduction from baseline for CBT. CBT = cognitive and behaviour therapy, S = scenario.}\label{tab:sim_scenarios}}
\end{table}

The interpretations of the eight scenarios are as follows~\cite{Lu2022}:
\begin{itemize}
\item {Scenario 1}: CBT has no effect
\item {Scenario 2}: CBT uniformly improves all three outcomes.
\item {Scenario 3}: CBT improves fatigue only (ranked first by 59\% of patients).
\item {Scenario 4}: CBT improves depression only (ranked first by 12\% of patients).
\item {Scenario 5:} CBT improves fatigue by effect size 1, depression by effect size 0.5 and has no effect for pain.
\item {Scenario 6}: CBT improves the top-ranked (= patient-selected) outcome by effect size 1, second-ranked outcome by effect size 0.5 and has no effect for the bottom-ranked outcome.
\item {Scenario 7}: CBT improves the bottom-ranked outcome by effect size 1, second-ranked outcome by effect size 0.5 and has no effect for the top-ranked (= patient-selected) outcome.
\item {Scenario 8}: CBT improves pain by effect size 1, \textit{worsens} fatigue by effect size 1 and has no effect on depression.
\end{itemize}
Note that scenarios 6 and 7 have a treatment effect by preference strata interaction. This could be caused by patients ranking the outcomes according to their baseline severity. Scenario~6 then corresponds to where CBT has more room for improvement on the more severe outcomes, while scenario~7 corresponds to where the more severe outcomes are harder to treat and less responsive to CBT.

As well as the `unequal' patient preferences \[(p_{123}, p_{132}, p_{213}, p_{231}, p_{312}, p_{321}) = (0.42, 0.17, 0.24, 0.05, 0.08, 0.04) \] elicited from the patient survey as described above, we also consider a hypothetical equal patient preference \[(p_{123}, p_{132}, p_{213}, p_{231}, p_{312}, p_{321}) = (1/6, 1/6, 1/6, 1/6, 1/6, 1/6)\] as a sensitivity analysis. \\

\noindent \textbf{Estimands or Other Targets}: We investigate the estimated power (type~I error rate under scenario~1) across the presented scenarios. \\

\noindent \textbf{Methods:} Each simulated dataset is analysed using the endpoints and corresponding analysis approaches described in Section~\ref{sec:methods}. \\

\noindent \textbf{Performance measures:} We assess the type~I error rate and power for each method.  The power (type~I error rate under scenario~1) in scenario~$c$ for method~$m$ is:
\[ \text{Power}_c = \frac{1}{n_{sim}} \sum_{i=1}^{n_{sim}} 1\{p_{i,m} \leq \alpha_{c,m}\} \]
where $n_{sim}$ is the number of simulation replicates, $p_{i,m}$ is the $p$-value for method~$m$ and simulation replicate~$i$, $\alpha_{1,m} = 0.05$ and is otherwise is method-dependent (see below). For each simulation scenario, we use $n_{sim} = 10^4$ so that the Monte Carlo standard error is less than 0.5\% in absolute terms.

\subsection{Results}
\label{subsec:results}

\noindent \textbf{Unequal preferences}

\noindent Table~\ref{tab:unequal_pref} show the power (type~I error rate for scenario 1) of the different methods under unequal patient preferences.

\begin{table}[ht]
\centering
\small
\begin{tabular}{lllllllll}
  \toprule
\textbf{Method} & \textbf{S1} & \textbf{S2} & \textbf{S3} & \textbf{S4} & \textbf{S5} & \textbf{S6} & \textbf{S7} & \textbf{S8} \\ 
  \midrule
  UV1& 4.7 & 98.7 & 98.7 & 4.7 & 98.7 & 97.4 & 7.7 & 0.0 \\ 
  UV2 & 5.0 & 98.4 & 5.0 & 5.0 & 5.0 & 45.0 & 70.1 & 98.4 \\ 
  UV3 & 4.9 & 98.5 & 4.9 & 98.5 & 60.0 & 13.6 & 94.3 & 4.9 \\ \midrule
  Composite DOOR & 6.0 & 99.8 & 59.9 & 26.4 & 76.6 & 80.9 & 59.2 & 1.8 \\ 
  WWP & 6.7 & 98.0 & 65.6 & 7.9 & 71.2 & 78.4 & 25.5 & 0.4 \\ 
  \midrule
  Mean patient-selected outcome & 4.8 & 98.7 & 68.5 & 8.8 & 74.8 & 81.6 & 28.0 & 0.4 \\ 
  Proportion test patient-selected  & 2.8 & 91.4 & 54.4 & 8.0 & 59.6 & 66.2 & 23.0 & 4.0 \\ 
   \bottomrule
\end{tabular}
\caption{Power (\%) under unequal patient preferences, based on $10^4$ replicates for each scenario. UV = univariate, DOOR = desirability of outcome ranking, WWP = weighted winning probabilities, S = scenario. \label{tab:unequal_pref}}
\end{table}

Starting with the null scenario (scenario~1), the type~I error rates of the univariate methods and the mean patient-selected outcome are controlled at the target 5\% level (within Monte Carlo simulation error), as would be expected. In contrast, composite DOOR and WWP have inflated type~I error rates of 6\%--7\%, due to ties in the \textit{U}-statistics and relatively small sample sizes~\cite{Lu2022}. Meanwhile, the proportion test has a deflated type~I error rate of approximately 3\%, due to the discrete nature of the test. In order to make a fair comparison of the power of the different methods for scenarios 2--8, we recalibrate the $p$-value threshold for composite DOOR, WWP and the proportion test so that the type~I error rate is controlled at~5\% or just below. 

For scenario~2, where the treatment improves all outcomes uniformly, patient preferences are not important in terms of power. Indeed, all methods have very high power (of between 98\%--100\%), except for the proportion test which has a power of 91\% that reflects the loss of efficiency due to dichotomisation.

In scenarios 3,4,5 and 8, the univariate approach has the highest power if the `correct' outcome (i.e., the outcome with a treatment effect of~1) is selected but otherwise has the lowest power if the `null' outcome (i.e., the outcome with a treatment effect of~0) is selected. The correct univariate approach can even have the highest power in scenarios 6 and 7 where there is a treatment by patient preference interaction, but this reflects (in scenario~6) how fatigue is ranked first or second by 91\% of patients and (in scenario~7) how depression is ranked last or second last by 88\% of patients.

Apart from the correct univariate approach,
composite DOOR has the highest power in scenarios 4 and 7, with substantially higher power than the mean-patient selected outcome. However, the mean patient-selected outcome has a substantially higher power than composite DOOR in scenario 3, and approximately the same power in scenarios 5 and 6, showing that it can be competitive with composite DOOR in scenarios where the (largest) treatment effect is seen in the outcome that a majority of patients deem the most important. The mean patient-selected outcome uniformly improves power slightly compared to WWP across scenarios 2--7. 

As would be expected, the proportion test has the lowest power (apart from the univariate approaches) for scenarios 2--7. Finally, for scenario~8, all methods apart from UV2 have very low power ($<5\%$) reflecting how the treatment leads to a worsening of fatigue scores. 

In terms of varying the correlation structure (Tables~\ref{tab:unequal_pref_low_corr} and~\ref{tab:unequal_pref_high_corr} in the Appendix), the univariate methods and patient-selected / top-ranked methods (WWP, mean patient-selected outcome and proportion test) are essentially unaffected (up to Monte Carlo error) as each outcome contributes to the test statistic independently from one another. In contrast, high correlation leads to power decreases (and low correlation leads to power increases) for composite DOOR.
Intuitively, this is because lower correlation means that each outcome contributes a greater amount of additional information that leads to increase power for composite DOOR.\\

\noindent \textbf{Equal preferences}

Table~\ref{tab:equal_pref} show the power (type~I error rate for scenario 1) of the different methods under the hypothetical scenario of completely equal patient preferences. The main difference with the unequal preferences setting is that composite DOOR has the highest power (apart from the `correct' univariate approach) for scenarios 3--7, and substantially outperforms the mean patient-selected outcome. This is not surprising given that only 1/3 of patients select each outcome as the top-ranked. However, the mean patient-selected outcome continues to uniformly improve power slightly compared to WWP across scenarios 2--7. Results for varying correlation structures can be found in Tables~\ref{tab:equal_pref_low_corr} and~\ref{tab:equal_pref_high_corr} in the Appendix, where again high correlation leads to power decreases (and low correlation leads to power increases) for composite DOOR. 

\begin{table}[ht!]
\centering
\small
\begin{tabular}{lllllllll}
  \toprule
\textbf{Method} & \textbf{S1} & \textbf{S2} & \textbf{S3} & \textbf{S4} & \textbf{S5} & \textbf{S6} & \textbf{S7} & \textbf{S8} \\ 
  \midrule
UV1 & 4.5 & 98.4 & 98.4 & 4.5 & 98.4 & 93.2 & 14.5 & 0.0 \\ 
  UV2 & 5.0 & 98.6 & 5.0 & 5.0 & 5.0 & 57.1 & 57.0 & 98.6 \\ 
  UV3 & 4.7 & 98.3 & 4.7 & 98.3 & 60.1 & 14.9 & 93.4 & 4.7 \\ \midrule
  Composite DOOR & 5.5 & 99.8 & 40.6 & 48.6 & 70.5 & 73.9 & 71.2 & 5.3 \\ 
  WWP & 6.6 & 97.9 & 29.7 & 30.9 & 54.3 & 65.1 & 42.4 & 4.2 \\ 
  \midrule
  Mean patient-selected outcome  & 4.8 & 98.5 & 32.7 & 31.4 & 57.3 & 68.8 & 43.8 & 3.7 \\ 
  Proportion test patient-selected  & 2.5 & 91.4 & 26.0 & 26.1 & 44.4 & 54.1 & 34.7 & 11.0 \\ 
   \bottomrule
\end{tabular}
\caption{Power (\%) under equal patient preferences, based on $10^4$ replicates for each scenario. UV = univariate, DOOR = desirability of outcome ranking, WWP = weighted winning probabilities, S = scenario. \label{tab:equal_pref}}
\end{table}

\section{Discussion}
\label{sec:discuss}


The simulation results in Section~\ref{sec:sim} illustrate how in settings where the majority of patients have the same top-ranked (i.e., selected) outcome, and there is medium or high correlation between outcomes, the mean patient-selected outcome can be competitive with composite DOOR in terms of power. Arguably, a medium (or even high) correlation between outcomes is a more plausible assumption to make in many disease settings. Purely from the perspective of statistical power, the distribution of patient preferences can be used to indicate \textit{a-priori} whether composite DOOR is preferable. The results also show that the mean patient-selected outcome has a uniformly higher power than WWP, and so on that basis the former is to be preferred. Finally, the proportion test tends to have low power across the simulation scenarios, and so cannot be recommended.

However, there are several other important considerations to take into account when making these comparisons. Firstly, composite DOOR relies on a meaningful preference ranking being reliably elicited from each patient. Due to the categorical preference ranking of outcomes, an implicit assumption is that there is an equal `distance' in terms of importance between each ranking step, which may not be true in practice. The use of patient-selected outcomes (and WWP) obviously removes the need for this assumption, and it is likely to be easier to elicit the single most important outcome compared with eliciting the full patient preference ranking.

At the same time, by ignoring the outcomes that are not the top-ranked/selected one, the use of patient-selected outcomes could miss meaningful differences in treatment effects, for example where multiple outcomes are seen as  quite similar in importance by patients. However, the flip side of this is that the increased power of composite DOOR can be driven by improvements on the outcome that (at least a majority of) patients rank as the least important, such as in Scenarios~4 and~7 in the simulation study. It is not clear that declaring a treatment as `successful' even if it does nothing to improve the most important outcomes for each patient (like in Scenario~7) would be acceptable to patients. 
These considerations underscore the potential advantage in further developing composite DOOR so that it can handle a continuous patient preference for outcomes that can be measured by a numerical weight for each outcome, although again this needs to be balanced with the difficulties of reliably eliciting such measures in practice.

Composite DOOR, WWP and the proportion test for the patient-selected outcome all explicitly incorporate the MCID in their calculations, unlike the mean patient-selected outcome. This is motivated (for composite DOOR and WWP) by requiring a `win' to have clinical relevance. The use of such `margins' like the MCID in DOOR calculations is debatable on statistical grounds given that it is rank-based test~\cite{Pocock2024}. For completeness, we present results without a margin (i.e.\ setting MCID = 0 for all outcomes) in Section~\ref{Asubsec:zero_margin}. These results show that the power of composite DOOR showed substantial increases under unequal preferences for Scenarios 2,3,5 and 7 (with power decreases for scenarios 4 and 6), but similar or decreased power under equal preferences. Meanwhile, the power of WWP and the proportion test slightly decreased.

Composite DOOR and WWP have the advantage of being non-parametric and hence can be straightforwardly applied to different types of outcomes. In contrast, the analysis methods for the patient-selected outcomes proposed in this paper have been parametric and assume normality, and hence would need to be adapted for other outcome types. The theoretical validity of the mean patient-selected outcome analysis (in terms of type~I error rate control) also relies on the use of a stratified randomisation scheme, which may not always be possible or straightforward in practice. The non-parametric approach of composite DOOR and WWP does come at a cost though in terms of type~I error rate inflation, at least when using the asymptotic normal approximation, which needs to be accounted for carefully in the analysis.

Finally, the use of patient-selected outcomes has demonstrated regulatory acceptability, as shown by the FDA guidance on migraine trials, and there is precedence to build on to apply such outcomes much more generally from a regulatory viewpoint. Another consideration is the interpretability of the different outcomes. The patient-selected outcomes have a much simpler interpretation than composite DOOR or WWP, which is particularly important when engaging with patients and the public.

We believe there is much scope for the further development, evaluation and practical use of patient-selected outcomes in clinical trials, particularly for medical conditions that affect multiple symptom domains. The two analysis methods for the patient-selected outcome proposed in this paper are relatively simple, and as future work it would be useful to consider more sophisticated analysis strategies to reduce the observed power gap with composite DOOR in some scenarios. For example, joint modelling of the outcomes would allow the borrowing of information and hence increased power. Similarly, the development of analysis strategies for different types of outcomes would be an important step. Finally, the incorporation of patient covariates into the analysis of patient-selected outcomes also requires further research.

\bibliography{patient_rank}

\clearpage
\section*{Appendix}
\appendix
\counterwithin{figure}{section}

\section{Composite DOOR algorithm}
\label{Asec:DOOR}

Let $S_{i_k}(r) = \{ l : R_{i,l} \leq r \}$ denote the index set for outcomes that are ranked as the top~$r$ most important by the $i$th patient on treatment arm~$k$. The composite DOOR algorithm is as follows for each pair $(i_0, i_1) \in \{1, \ldots, n_0\} \times \{1, \ldots, n_1\}$:
\begin{enumerate}
\setcounter{enumi}{-1}
\item Set $r =0$
\item Set $r = r+1$
\item Does $S_{i_0}(r) = S_{i_1}(r)$? If no, go to step 1. If yes, go to step 3. 
\item If $\displaystyle \max_{j \in S_{i_0}(r)} Z_{i_0, i_1}(j) = 1$ and $\displaystyle \min_{j \in S_{i_0}(r)} Z_{i_0, i_1}(j) >0$, set $Z_{i_0, i_1} = 1$ and end the composite DOOR calculation. Otherwise, go to step 4.
\item If $\displaystyle \max_{j \in S_{i_0}(r)} Z_{i_0, i_1}(j) < 1$ and $\displaystyle \min_{j \in S_{i_0}(r)} Z_{i_0, i_1}(j) =0$, set $Z_{i_0, i_1} = 0$ and end the composite DOOR calculation. Otherwise, go to step 5.
\item If $\displaystyle \max_{j \in S_{i_0}(r)} Z_{i_0, i_1}(j) = 1$ and $\displaystyle \min_{j \in S_{i_0}(r)} Z_{i_0, i_1}(j) = 0$, or $r = m$, set $Z_{i_0, i_1} = 0.5$ and end the composite DOOR calculation. Otherwise, go to step 1. \\
\end{enumerate}

The estimated standard deviation $\hat{\sigma}$ used for hypothesis testing is given by $\sqrt{\widehat{\text{var}(\hat{\theta})}}$, where
\begin{align*}
\widehat{\text{var}(\hat{\theta})} = & \; \frac{1}{n_0 n_1}\left( \frac{1}{n_0 n_1}\sum_{i_0 =1}^{n_0} \sum_{i_1 =1}^{n_1} Z_{i_0, i_1}^2 +  \frac{1}{n_0 n_1}\sum_{i_0 =1}^{n_0} \sum_{i_1 =1}^{n_1} \sum_{i_1' = 1, i_1' \neq i_1}^{n_1} \! \! \! Z_{i_0, i_1} Z_{i_0, i_1'} \, +  \right. \\
& \; \left.  \frac{1}{n_0 n_1}\sum_{i_0 =1}^{n_0} \sum_{i_1 =1}^{n_1} \sum_{i_0' = 1, i_0' \neq i_0}^{n_0} \!\! \! Z_{i_0, i_1} Z_{i_0', i_1}\, - (n -1) \hat{\theta}^2 \right). \\
\end{align*}

\section{Top-ranked weighted winning probability}
\label{Asec:WWT}

Let the vector of within-stratum winning probabilities be denoted $\hat{\boldsymbol{\theta}} = [\hat{\theta}_1, \dots, \hat{\theta}_m]^T$ and the vector of the sample proportions of the proportion of patients in each stratum be denoted $\hat{\boldsymbol{p}} = [\hat{p}_1, \dots, \hat{p}_m]^T$.

The estimated standard deviation $\hat{\sigma}_{WT}$ used for hypothesis testing is given by $\sqrt{\widehat{\text{var}(\hat{\theta}_{WT})}}$, where
\begin{equation*}
    \widehat{\text{var}(\hat{\theta}_{WT})} = \hat{\boldsymbol{\theta}}^T \Sigma_{\hat{\boldsymbol{p}}} \hat{\boldsymbol{\theta}} + \hat{\boldsymbol{p}}^T \text{diag}(\text{Var}(\hat{\boldsymbol{\theta}})) \hat{\boldsymbol{p}}
\end{equation*}
Here $\text{var}(\hat{\boldsymbol{\theta}}) = (\text{var}(\hat{\theta}_1), \cdots, \text{var}(\hat{\theta}_m))$ and $\Sigma_{\hat{\boldsymbol{p}}} = \frac{1}{n} \left( \text{diag}(\hat{\boldsymbol{p}}) - \hat{\boldsymbol{p}}\hat{\boldsymbol{p}}^T \right)$.

\section{Additional simulation results}
\label{Asec:sim}

\setcounter{table}{0}
\renewcommand{\thetable}{C\arabic{table}}

\subsection{Results for low correlation}

\begin{table}[H]
\centering
\small
\begin{tabular}{lllllllll}
  \toprule
\textbf{Method} & \textbf{S1} & \textbf{S2} & \textbf{S3} & \textbf{S4} & \textbf{S5} & \textbf{S6} & \textbf{S7} & \textbf{S8} \\ 
  \midrule
UV1 & 4.8 & 98.5 & 98.5 & 4.8 & 98.5 & 97.3 & 7.5 & 0.0 \\ 
  UV2 & 5.0 & 98.3 & 5.0 & 5.0 & 5.0 & 45.1 & 69.9 & 98.3 \\ 
  UV3 & 4.9 & 98.3 & 4.9 & 98.3 & 60.0 & 13.8 & 94.2 & 4.9 \\ \midrule
  Composite DOOR & 5.9 & 100.0 & 72.0 & 26.1 & 86.2 & 90.2 & 63.8 & 1.2 \\ 
  WWP & 6.6 & 97.9 & 65.7 & 7.9 & 71.7 & 79.0 & 25.3 & 0.4 \\ 
  \midrule
  Mean patient-selected outcome & 4.9 & 98.8 & 68.8 & 8.8 & 74.9 & 81.8 & 28.4 & 0.4 \\ 
  Proportion test patient-selected  & 2.8 & 91.5 & 54.5 & 8.2 & 59.6 & 65.9 & 23.2 & 3.8 \\ 
   \bottomrule
\end{tabular}
\caption{Power (\%) under unequal patient preferences and low correlation, based on $10^4$ replicates for each scenario. UV = univariate, DOOR = desirability of outcome ranking, WWP = weighted winning probabilities, S = scenario. \label{tab:unequal_pref_low_corr}}
\end{table}

\begin{table}[H]
\centering
\small
\begin{tabular}{lllllllll}
  \toprule
\textbf{Method} & \textbf{S1} & \textbf{S2} & \textbf{S3} & \textbf{S4} & \textbf{S5} & \textbf{S6} & \textbf{S7} & \textbf{S8} \\ 
  \midrule
UV1 & 4.6 & 98.3 & 98.3 & 4.6 & 98.3 & 93.5 & 15.1 & 0.0 \\ 
  UV2 & 5.1 & 98.6 & 5.1 & 5.1 & 5.1 & 57.1 & 57.0 & 98.6 \\ 
  UV3 & 4.9 & 98.4 & 4.9 & 98.4 & 60.0 & 15.1 & 93.2 & 4.9 \\ \midrule
  Composite DOOR. & 5.5 & 100.0 & 48.4 & 55.7 & 80.6 & 83.5 & 79.1 & 5.0 \\ 
  WWP & 6.7 & 97.9 & 30.0 & 30.8 & 54.4 & 65.4 & 42.0 & 4.4 \\ 
  \midrule
   Mean patient-selected outcome  & 4.6 & 98.4 & 32.9 & 31.6 & 57.1 & 68.7 & 43.6 & 3.6 \\ 
  Proportion test patient-selected & 2.6 & 91.3 & 26.4 & 26.6 & 44.1 & 54.2 & 35.3 & 11.3 \\ 
   \bottomrule
\end{tabular}
\caption{Power (\%) under equal patient preferences and low correlation, based on $10^4$ replicates for each scenario. UV = univariate, DOOR = desirability of outcome ranking, WWP = weighted winning probabilities, S = scenario. \label{tab:equal_pref_low_corr}}
\end{table}

\subsection{Results for high correlation}

\begin{table}[H]
\centering
\small
\begin{tabular}{lllllllll}
  \toprule
\textbf{Method} & \textbf{S1} & \textbf{S2} & \textbf{S3} & \textbf{S4} & \textbf{S5} & \textbf{S6} & \textbf{S7} & \textbf{S8} \\ 
  \midrule
  UV1 & 5.0 & 98.6 & 98.6 & 5.0 & 98.6 & 97.4 & 7.6 & 0.0 \\ 
  UV2 & 5.1 & 98.5 & 5.1 & 5.1 & 5.1 & 45.1 & 70.3 & 98.5 \\ 
  UV3 & 4.8 & 98.6 & 4.8 & 98.6 & 60.1 & 13.6 & 94.3 & 4.8 \\ \midrule
  Composite DOOR & 5.9 & 99.2 & 50.3 & 30.1 & 67.9 & 70.5 & 58.0 & 2.2 \\ 
  WWP & 6.9 & 97.9 & 65.2 & 7.8 & 71.0 & 78.3 & 25.3 & 0.4 \\ 
  \midrule
  Mean patient-selected outcome & 5.0 & 98.8 & 68.3 & 9.0 & 74.4 & 81.4 & 28.2 & 0.4 \\ 
  Proportion test patient-selected & 2.5 & 91.7 & 54.2 & 7.8 & 59.4 & 66.2 & 23.1 & 4.1 \\ 
   \bottomrule
\end{tabular}
\caption{Power (\%) under unequal patient preferences and high correlation, based on $10^4$ replicates for each scenario. UV = univariate, DOOR = desirability of outcome ranking, WWP = weighted winning probabilities, S = scenario. \label{tab:unequal_pref_high_corr}}
\end{table}

\begin{table}[H]
\centering
\small
\begin{tabular}{lllllllll}
  \toprule
\textbf{Method} & \textbf{S1} & \textbf{S2} & \textbf{S3} & \textbf{S4} & \textbf{S5} & \textbf{S6} & \textbf{S7} & \textbf{S8} \\ 
  \midrule
  UV1 & 4.5 & 98.4 & 98.4 & 4.5 & 98.4 & 93.1 & 14.8 & 0.0 \\ 
  UV2 & 4.9 & 98.5 & 4.9 & 4.9 & 4.9 & 57.1 & 57.4 & 98.5 \\ 
  UV3 & 4.5 & 98.5 & 4.5 & 98.5 & 60.1 & 14.9 & 93.3 & 4.5 \\ \midrule
 Composite DOOR & 5.6 & 99.1 & 36.7 & 46.8 & 63.9 & 65.0 & 66.2 & 5.1 \\ 
  WWP & 6.5 & 97.9 & 29.7 & 30.9 & 54.7 & 65.0 & 42.0 & 4.2 \\ 
  \midrule
  Mean patient-selected outcome & 4.8 & 98.5 & 32.3 & 31.3 & 57.3 & 68.7 & 43.4 & 3.7 \\ 
  Proportion test patient-selected & 2.4 & 91.5 & 26.1 & 26.1 & 44.3 & 54.2 & 34.6 & 10.8 \\ 
  \bottomrule
\end{tabular}
\caption{Power (\%) under equal patient preferences and high correlation between outcomes, based on $10^4$ replicates for each scenario. UV = univariate, DOOR = desirability of outcome ranking, WWP = weighted winning probabilities, S = scenario. \label{tab:equal_pref_high_corr}}
\end{table}

\clearpage
\subsection{Results under zero margin}
\label{Asubsec:zero_margin}

\noindent \textbf{Unequal preferences}

\begin{table}[H]
\centering
\small
\begin{tabular}{lllllllll}
  \toprule
\textbf{Method} & \textbf{S1} & \textbf{S2} & \textbf{S3} & \textbf{S4} & \textbf{S5} & \textbf{S6} & \textbf{S7} & \textbf{S8} \\ 
  \midrule
  Composite DOOR & 6.0 & 99.5 & 74.6 & 11.6 & 81.7 & 87.5 & 34.9 & 0.4 \\ 
  WWP & 6.5 & 97.6 & 64.7 & 7.6 & 70.4 & 77.7 & 24.4 & 0.5 \\ 
  \midrule
  Proportion test patient selected & 2.6 & 89.4 & 46.4 & 7.3 & 52.3 & 59.1 & 18.9 & 0.6 \\ 
   \bottomrule
\end{tabular}
\caption{Power (\%) under unequal patient preferences and zero margin, based on $10^4$ replicates for each scenario. DOOR = desirability of outcome ranking, WWP = weighted winning probabilities, S = scenario. \label{tab:unequal_pref_margin0}}
\end{table}

\noindent \textbf{Equal preferences}

\begin{table}[H]
\centering
\small
\begin{tabular}{lllllllll}
  \toprule
\textbf{Method} & \textbf{S1} & \textbf{S2} & \textbf{S3} & \textbf{S4} & \textbf{S5} & \textbf{S6} & \textbf{S7} & \textbf{S8} \\ 
  \midrule
  Composite DOOR & 4.7 & 99.6 & 39.5 & 39.6 & 68.2 & 75.4 & 61.7 & 4.7 \\ 
  WWP & 4.5 & 97.6 & 29.0 & 29.3 & 53.1 & 64.0 & 40.3 & 4.1 \\ 
  \midrule
  Proportion test patient selected & 4.6 & 89.4 & 20.7 & 20.4 & 38.6 & 46.2 & 28.7 & 4.1 \\ 
   \bottomrule
\end{tabular}
\caption{Power (\%) under equal patient preferences and zero margin, based on $10^4$ replicates for each scenario. DOOR = desirability of outcome ranking, WWP = weighted winning probabilities, S = scenario. \label{tab:equal_pref_margin0}}
\end{table}

\noindent \textbf{Unequal preferences (low correlation)}

\begin{table}[H]
\centering
\small
\begin{tabular}{lllllllll}
  \toprule
\textbf{Method} & \textbf{S1} & \textbf{S2} & \textbf{S3} & \textbf{S4} & \textbf{S5} & \textbf{S6} & \textbf{S7} & \textbf{S8} \\ 
  \midrule
  Composite DOOR & 6.1 & 99.8 & 80.4 & 11.9 & 86.6 & 91.9 & 37.3 & 0.3 \\ 
  WWP & 6.8 & 97.7 & 65.0 & 7.7 & 70.5 & 77.9 & 24.4 & 0.4 \\ 
  \midrule
  Proportion test patient selected & 2.7 & 89.4 & 46.7 & 7.6 & 52.6 & 59.0 & 18.8 & 0.6 \\ 
   \bottomrule
\end{tabular}
\caption{Power (\%) under unequal patient preferences, low correlation between outcomes and zero margin, based on $10^4$ replicates for each scenario. DOOR = desirability of outcome ranking, WWP = weighted winning probabilities, S = scenario. \label{tab:unequal_pref_margin0_low_corr}}
\end{table}

\clearpage
\noindent \textbf{Equal preferences (low correlation)}

\begin{table}[H]
\centering
\small
\begin{tabular}{lllllllll}
  \toprule
\textbf{Method} & \textbf{S1} & \textbf{S2} & \textbf{S3} & \textbf{S4} & \textbf{S5} & \textbf{S6} & \textbf{S7} & \textbf{S8} \\ 
  \midrule
 Composite DOOR & 5.9 & 99.9 & 44.6 & 44.4 & 75.3 & 82.4 & 68.5 & 4.7 \\ 
  WWP & 6.8 & 97.8 & 29.5 & 29.4 & 53.2 & 64.2 & 40.3 & 4.3 \\ 
  \midrule
 Proportion test patient selected & 2.7 & 88.9 & 20.7 & 20.8 & 39.0 & 47.2 & 29.0 & 3.9 \\ 
   \bottomrule
\end{tabular}
\caption{Power (\%) under equal patient preferences, low correlation between outcomes and zero margin, based on $10^4$ replicates for each scenario. DOOR = desirability of outcome ranking, WWP = weighted winning probabilities, S = scenario. \label{tab:equal_pref_margin0_low_corr}}
\end{table}

\noindent \textbf{Unequal preferences (high correlation)}

\begin{table}[H]
\centering
\small
\begin{tabular}{lllllllll}
  \toprule
\textbf{Method} & \textbf{S1} & \textbf{S2} & \textbf{S3} & \textbf{S4} & \textbf{S5} & \textbf{S6} & \textbf{S7} & \textbf{S8} \\ 
  \midrule
  Composite DOOR & 6.0 & 99.1 & 70.8 & 11.7 & 77.9 & 83.8 & 33.0 & 0.5 \\ 
  WWP & 6.6 & 97.6 & 64.6 & 7.7 & 70.0 & 77.4 & 24.0 & 0.5 \\ 
  \midrule
  Proportion test patient selected & 2.8 & 89.0 & 46.3 & 7.2 & 52.0 & 58.7 & 18.9 & 0.6 \\ 
   \bottomrule
\end{tabular}
\caption{Power (\%) under unequal patient preferences, high correlation between outcomes and zero margin, based on $10^4$ replicates for each scenario. DOOR = desirability of outcome ranking, WWP = weighted winning probabilities, S = scenario. \label{tab:unequal_pref_margin0_high_corr}}
\end{table}

\noindent \textbf{Equal preferences (high correlation)}

\begin{table}[H]
\centering
\small
\begin{tabular}{lllllllll}
  \toprule
\textbf{Method} & \textbf{S1} & \textbf{S2} & \textbf{S3} & \textbf{S4} & \textbf{S5} & \textbf{S6} & \textbf{S7} & \textbf{S8} \\ 
  \midrule
  Composite DOOR & 5.6 & 99.1 & 37.4 & 37.0 & 63.1 & 70.3 & 56.5 & 4.5 \\ 
  WWP& 6.8 & 97.5 & 29.0 & 28.8 & 53.3 & 64.3 & 40.2 & 4.0 \\ 
  \midrule
  Proportion test patient selected & 2.6 & 89.0 & 20.8 & 20.4 & 38.1 & 46.3 & 28.6 & 3.7 \\ 
   \bottomrule
\end{tabular}
\caption{Power (\%) under equal patient preferences, high correlation between outcomes and zero margin, based on $10^4$ replicates for each scenario. DOOR = desirability of outcome ranking, WWP = weighted winning probabilities, S = scenario. \label{tab:equal_pref_margin0_high_corr}}
\end{table}

\end{document}